\documentclass[prl,twocolumn,superscriptaddress,showpacs]{revtex4}

\usepackage{amsfonts}
\usepackage{amssymb}
\usepackage{amsmath}
\usepackage{hyperref}
\usepackage{graphicx}
\usepackage{longtable}

\newcommand{\calC}{\mathcal{C}}
\newcommand{\calI}{\mathcal{I}}

\newtheorem{definition}{Definition}

\begin{document}
\title{Ramsey numbers and adiabatic quantum computing}

\date{\today}

\author{Frank Gaitan}
\affiliation{Laboratory for Physical Sciences, 8050 Greenmead Dr, 
College Park, MD 20740}

\author{Lane Clark}
\affiliation{Department of Mathematics, Southern Illinois University, 
Carbondale, IL 62901-4401}

\begin{abstract}
The graph-theoretic Ramsey numbers are notoriously difficult to calculate. In
fact, for the two-color Ramsey numbers $R(m,n)$ with $m,n\geq 3$, only nine 
are currently known. We present a quantum algorithm for the computation of the 
Ramsey numbers $R(m,n)$. We show how the computation of $R(m,n)$ can be mapped
to a combinatorial optimization problem whose solution can be found using adiabatic
quantum evolution. We numerically simulate this adiabatic quantum algorithm and 
show that it correctly determines the Ramsey numbers $R(3,3)$ and $R(2,s)$ for 
$5\leq s\leq 7$. We then discuss the algorithm's experimental implementation, 
and close by showing that Ramsey number computation belongs to the quantum 
complexity class QMA. 
\end{abstract}


\maketitle

In an arbitrary party of $N$ people one might ask whether there is a group 
of $m$ people who are all mutually acquainted, or a group of $n$ people who
are all mutual strangers. Using Ramsey theory \cite{grah,hand}, it can be shown
that a threshold value $R(m,n)$ exists for the party size $N$ so that 
when $N\geq R(m,n)$, all parties of $N$ people will either contain $m$ mutual
acquaintances, or $n$ mutual strangers. The threshold value $R(m,n)$ is an
example of a two-color Ramsey number. Other types of Ramsey numbers exist, 
though we will focus on two color Ramsey numbers in this paper.

One can represent the $N$-person party problem by an $N$-vertex graph.
Here each person is associated with a vertex, and an edge is drawn between 
a pair of vertices only when the corresponding people know each other. In 
the case where $m$ people are mutual acquaintances, there will be an edge 
connecting any pair of the $m$ corresponding vertices. Similarly, if 
$n$ people are mutual strangers, there will be no edge between any of the 
$n$ corresponding vertices. In the language of graph theory \cite{boll}, the 
$m$ vertices form an $m$-clique, and the $n$ vertices form an 
$n$-independent set. The party problem is now a statement in graph theory: 
if $N\geq R(m,n)$, every graph with $N$ vertices will contain either an 
$m$-clique, or an $n$-independent set. Ramsey numbers can also be introduced 
using colorings of complete graphs, and $R(m,n)$ corresponds to the case where 
only two colors are used.

Ramsey theory has found applications in mathematics, information theory, and
theoretical computer science \cite{rosta}. An application of
fundamental significance appears in the Paris-Harrington (PH) theorem of 
mathematical logic \cite{PHthm} which established that a particular statement 
in Ramsey theory related to graph colorings and natural numbers is true, though 
unprovable within the axioms of Peano arithmetic. Such statements are known 
to exist as a consequence of Godel's incompleteness theorem, though the PH 
theorem provided the first natural example. Deep connections have also been 
shown to exist between Ramsey theory, topological dynamics, and ergodic theory
\cite{furst}.

Ramsey numbers grow extremely quickly and so are notoriously difficult to 
calculate. In fact, for two color Ramsey numbers $R(m,n)$ with $m,n\geq 3$, 
only nine are presently known\cite{boll}. To check whether $N\stackrel{?}{=}
R(m,n)$ requires examining \textit{all\/} $2^{N(N-1)/2}$ $N$-vertex graphs. 
The number of graphs to be checked thus grows super-exponentially with $N$, 
and so the task quickly becomes intractable. Ketonen and Solovay \cite{Ket&Sol} 
have shown that this is the root cause for why the statement in the PH theorem
cannot be proved within Peano arithmetic. 

In this paper we: (i)~present a quantum algorithm for calculating Ramsey 
numbers based on adiabatic quantum evolution; (ii)~numerically simulate
the algorithm to verify that it correctly calculates small Ramsey numbers; 
(iii)~discuss its experimental implementation; and (iv)~show that Ramsey 
number computation belongs to the quantum complexity class \textit{QMA}.

\textbf{Optimization Problem:} We begin by establishing a 1-1 correspondence 
between the set of 
$N$-vertex graphs and binary strings of length $L=N(N-1)/2$. To each 
$N$-vertex graph $G$ there corresponds a unique adjacency matrix $A(G)$ 
which is an $N\times N$ symmetric matrix with vanishing diagonal matrix 
elements, and with off-diagonal element $a_{i,j}=1\, (0)$ when distinct 
vertices $i$ and $j$ are (are not) joined by an edge. It follows that $A(G)$ 
is determined by its lower triangular part. By concatenating column-wise 
the matrix elements $a_{i,j}$ appearing below the principal diagonal, we 
can construct a unique binary string $g(G)$ of length $L$ for each graph $G$:
\begin{equation}
g(G) \equiv a_{2,1}\cdots a_{N,1}\;a_{3,2}\cdots a_{N,2}\;\cdots \;a_{N,N-1}.
\label{121corr}
\end{equation}

Given the string $g(G)$, the following procedure determines the number of 
$m$-cliques in $G$. Choose $m$ vertices $S_{\alpha} = \{ v_{1},\ldots ,
v_{m}\}$ from the $N$ vertices of $G$ and form the product
$\calC_{\alpha} = \prod_{(v_{j},v_{k}\in S_{\alpha})}^{(j\neq k)} 
                                  a_{v_{j},v_{k}} $.
Note that $\calC_{\alpha}=1$ when $S_{\alpha}$ forms an $m$-clique; 
otherwise $\calC_{\alpha}=0$. Now repeat 
this procedure for all $\rho = C(N,m)$ 
ways of choosing $m$ vertices from $N$ vertices, and form the sum
$\calC (G) = \sum_{\alpha =1}^{\rho} \calC_{\alpha}$.
By construction, $\calC (G)$ equals the number of $m$-cliques contained in
$G$. A similar procedure determines the number of $n$-independent sets in
$G$. Briefly, choose $n$ vertices $T_{\alpha} = \{v_{1},\ldots ,v_{n}\}$ from 
the $N$ vertices in $G$, and form the product
$\calI_{\alpha} = \prod_{(v_{j},v_{k}\in T_{\alpha})}^{(j\neq k)}
                               \overline{a}_{v_{j},v_{k}}$, where 
$\overline{a}_{v_{j},v_{k}} = 1 -a_{v_{j},v_{k}}$. If the vertex set 
$T_{\alpha}$ forms an $n$-independent set, then $\calI_{\alpha} =1$; 
otherwise $\calI_{\alpha}= 0$. Repeat this for all $\nu = C(N,n)$ ways of 
choosing $n$ vertices from $N$ vertices, then form the sum 
$\calI (G) = \sum_{\alpha =1}^{\nu}\calI_{\alpha}$. By construction, 
$\calI (G)$ gives the number of $n$-independent sets contained in $G$. 
Finally, define
\begin{equation}
h(G) = \calC (G) + \calI (G) .
\label{grafcost}
\end{equation}
It follows from the above discussion that $h(G)$ is the total number of 
$m$-cliques and $n$-independent sets in $G$. Thus $h(G) \geq 0$ for all 
graphs $G$; and $h(G) = 0$ if and only if $G$ does not contain an $m$-clique 
or $n$-independent set. 

We can use $h(G)$ as the cost function for the following combinatorial 
optimization problem. For given integers ($N, m, n$), and 
with $h(G)$ defined as above, find an $N$-vertex graph $G_{\ast}$
that yields the global minimum of $h(G)$. Notice that if $N<R(m,n)$, the 
(global) minimum will be $h(G_{\ast})=0$ since Ramsey theory guarantees that 
a graph exists which has no $m$-clique or $n$-independent set. On the other 
hand, if $N\geq R(m,n)$, Ramsey theory guarantees $h(G_{\ast})>0$. If 
we begin with $N<R(m,n)$ and increment $N$ by $1$ until we first find 
$h(G_{\ast})> 0$, then the corresponding $N$ will be exactly $R(m,n)$.
We now show how this combinatorial optimization problem can be solved 
using adiabatic quantum evolution. 

\textbf{Quantum Algorithm:} The adiabatic quantum evolution (AQE) 
algorithm \cite{aqe} exploits the adiabatic dynamics of a quantum system to 
solve combinatorial optimization problems. The AQE algorithm uses the 
optimization problem cost function to define a problem Hamiltonian $H_{P}$ 
whose ground-state subspace encodes all problem solutions. The algorithm 
evolves the state of an $L$-qubit register from the ground-state of an initial 
Hamiltonian $H_{i}$ to the ground-state of $H_{P}$ with probability approaching 
$1$ in the adiabatic limit. An appropriate measurement at the end of the 
adiabatic evolution yields a solution of the optimization problem almost 
certainly. The time-dependent Hamiltonian $H(t)$ for global AQE is
\begin{equation}
H(t) = \left( 1 - \frac{t}{T}\right) H_{i} + \left(\frac{t}{T}\right) H_{P},
\label{aqeHam}
\end{equation}
where $T$ is the algorithm runtime, and adiabatic dynamics corresponds 
to $T\rightarrow \infty$.

To map the optimization problem associated with computing $R(m,n)$ onto
an adiabatic quantum computation, we begin with the 1-1 correspondence
between $N$-vertex graphs $G$ and length $L=N(N-1)/2$ binary
strings $g(G)$. From Eq.~(\ref{121corr}) we see that position along the string
is indexed by vertex pairs ($i,j$). We thus identify a qubit with each such pair 
($i,j$), and will thus need $L$ qubits. Defining the computational basis states
(CBS) to be the eigenstates of $\sigma_{z}^{0}\otimes\cdots\otimes
\sigma_{z}^{L-1}$, we identify the $2^{L}$ graph strings $g(G)$ with the 
$2^{L}$ CBS: 
$g(G)\rightarrow |g(G)\rangle$. The problem Hamiltonian $H_{P}$ is defined 
to be diagonal in the computational basis with eigenvalue $h(G)$ 
associated with eigenstate $|g(G)\rangle$:
\begin{equation}
H_{P}|g(G)\rangle = h(G)|g(G)\rangle .
\label{HPdef}
\end{equation}
Note that the ground-state energy of $H_{P}$ will be zero iff there is a graph 
with no $m$-cliques or $n$-independent sets. We give an operator 
expression for $H_{P}$ below. The initial Hamiltonian $H_{i}$ is chosen to be
\begin{equation}
H_{i} = \sum_{l=0}^{L-1}\frac{1}{2}\left( I^{l} - \sigma_{x}^{l}\right) ,
\label{Hidef}
\end{equation}
where $I^{l}$ and $\sigma_{x}^{l}$ are the identity and x-Pauli operator
for qubit $l$, respectively. The ground-state of $H_{i}$ 
is the easily constructed uniform superposition of CBS.

The quantum algorithm for computing $R(m,n)$ begins by setting $N$ equal
to a strict lower bound for $R(m,n)$ which can be found using the probabilistic
method \cite{spenc} or a table of two-color Ramsey numbers \cite{boll}.
The AQE algorithm is run on $L_{N}=N(N-1)/2$ qubits, and the energy $E$ is 
measured at the end of algorithm execution. In the \textit{adiabatic limit\/} 
the result will be $E=0$ since $N<R(m,n)$. The value of $N$ is now incremented 
$N\rightarrow N+1$, the AQE algorithm is re-run on $L_{N+1}$ qubits, and the 
energy $E$ measured at the end of algorithm execution. This process is repeated 
until $E>0$ first occurs, at which point the associated $N$ will be equal to 
$R(m,n)$. Note that any real application of AQE will only be approximately 
adiabatic. Thus the probability that the measured energy $E$ will be the 
ground-state energy will be $1-\epsilon$. In this case, the algorithm must be 
run $k\sim\mathcal{O}(\ln [1-\delta ]/\ln\epsilon)$ times so that, with 
probability $\delta > 1-\epsilon$, at least one of the measurement outcomes 
will be the true ground-state energy. We can make $\delta$ arbitrarily close 
to $1$ by choosing $k$ sufficiently large.

\textbf{Simulation Results:} To test the adiabatic quantum computation of 
$R(m,n)$, we numerically simulated the Schrodinger dynamics generated by 
the AQE Hamiltonian $H(t)$. Clearly, these simulations can only be run at 
finite values of $T$. As in Ref.~\cite{farhi}, we chose $T$ so that the 
algorithm success probability $P_{s}$ is large compared to the 
probability that a randomly chosen CBS will belong to the 
$D$-degenerate ground-state eigenspace of $H_{P}$ ($P_{s}\gg D/2^{L})$). 
Here $P_{s}$ is the probability that an energy 
measurement done at the final time $T$ will yield the ground-state energy 
$E_{gs}$ of $H_{P}$. Since a classical computer cannot efficiently simulate 
the dynamics of a quantum system, we can only obtain small Ramsey 
numbers. In this case, $H_{P}$ can be found by evaluating 
the cost function $h(G)$ using the procedure described above 
Eq.~(\ref{grafcost}). 

We simulated the AQE computation of $R(3,3)$ and $R(2,s)$ for $5\leq s\leq 7$. 
Straightforward arguments \cite{boll} give $R(3,3)=6$ and $R(2,s)=s$. 
We present our simulation results in Table~\ref{table1}. 
\begin{table*}[h,t,b]
\caption{\label{table1}Simulation results for Ramsey numbers $R(3,3)$ and
$R(2,s)$ for $5\leq s\leq 7$. Here $N$ is the number of graph vertices; 
$E_{gs}$ and $D$ are the ground-state energy and degeneracy,
respectively, for the problem Hamiltonian $H_{P}$; and $T$ and $P_{s}$ 
are, respectively, the algorithm runtime and success probability.}
\begin{ruledtabular}
\begin{tabular}{|c|cc|cc||c|cc|cc||c|cc|cc||c|cc|cc|}
\multicolumn{5}{|c||}{$\mathbf{R(2,5)}$} & 
  \multicolumn{5}{c||}{$\mathbf{R(2,6)}$} & 
    \multicolumn{5}{c||}{$\mathbf{R(3,3)}$} &
     \multicolumn{5}{c|}{$\mathbf{R(2,7)}$}\\\hline
$N$ & $E_{gs}$ & $D$ & $T$ & $P_{s}$ &
$N$ & $E_{gs}$ & $D$ & $T$ & $P_{s}$ &
$N$ & $E_{gs}$ & $D$ & $T$ & $P_{s}$ &
$N$ & $E_{gs}$ & $D$ & $T$ & $P_{s}$ \\\hline
$3$ & $0.0$ & $1$ & $5.0$ & $0.591$ &
$4$ & $0.0$ & $1$ & $5.0$ & $0.349$ &
$4$  & $0.0$ & $18$ & $5.0$ & $0.769$ &
$5$  & $0.0$ & $1$ & $8.0$ & $0.865$ \\
$4$ & $0.0$ & $1$ & $5.0$  & $0.349$ &
$5$ & $0.0$ & $1$ & $5.0$ & $0.173$ &
$5$ & $0.0$ & $12$ & $5.0$ & $0.194$ &
$6$ & $0.0$ & $1$ & $8.0$ & $0.805$ \\
$5$ & $1.0$ & $11$ & $5.0$ & $0.518$ &
$6$ & $1.0$ & $16$ & $5.0$ & $0.286$ &
$6$ & $2.0$ & $1760$ & $5.0$ & $0.693$ &
$7$ & $1.0$ & $22$ & $8.0$ & $0.938$ \\
\end{tabular}
\end{ruledtabular}
\end{table*}
 We see that for all $m,n$ considered, the threshold value $N_{t}$ where 
$E_{gs}>0$ first occurs is precisely at the Ramsey number: $N_{t}=R(m,n)$. 

For $R(2,s)$ and $N=s$, Table~\ref{table1} gives $E_{gs}=1$. For these cases,
graphs corresponding to ground-states of $H_{P}$ will thus contain either a 
single $s$-independent set or a single $2$-clique. There is only one $s$-vertex 
graph with an $s$-independent set, and there are $C(s,2)=s(s-1)/2$ graphs with 
one $2$-clique (viz.~edge). Thus the ground-state degeneracy $D=1+C(s,2)$, 
in agreement with the $R(2,s)$ degeneracies in Table~\ref{table1} for $N=s=
5,6,7$. For $R(3,3)$ and $N=6$, Table~\ref{table1} gives $E_{gs}=2$. Thus
graphs corresponding to ground-states are those with: (i)~two $3$-cliques; 
(ii)~two $3$-independent sets; or (iii)~one $3$-clique and one $3$-independent set. 
Ref.~\cite{good} derived the minimum number of $3$-cliques and $3$-independent 
sets that can be present in an $N$-vertex graph. This minimum is precisely our 
$E_{gs}$ for $R(3,3)$ and a given $N$. For $N=6$, the minimum value is $2$, 
in agreement with $E_{gs}=2$ in Table~\ref{table1}. We carried out both 
analytical \cite{lhc} and numerical counts of the ground-state graphs for 
$R(3,3)$ and $N=6$. Both approaches found $1760$ graphs giving a  
ground-state degeneracy $D=1760$. In all cases appearing in Table~\ref{table1},
the upward jump in $D$ seen upon reaching the Ramsey threshold $N=R(m,n)$ 
(from below) is responsible for the jump in the success probability $P_{s}$ 
also seen at this threshold.

Although we would like to have calculated larger Ramsey numbers, this was 
simply not practical. Note that the $N=7$ simulations use $L=21$ qubits. These
simulations are at the upper limit of $20$-$22$ qubits at which simulation 
of the full AQE Schrodinger dynamics is practical \cite{farhi,fg1,fg2}. The 
next smallest Ramsey number is $R(2,8)=8$ which requires a $28$ qubit 
simulation, well beyond what can be done practically.

\textbf{Experimental Implementation:} We begin by determining an operator
expression for the problem Hamiltonian $H_{P}$ which then fixes the AQE 
Hamiltonian $H(t)$ through Eqs.~(\ref{aqeHam}) and (\ref{Hidef}). Recall that 
the eigenvalue $h(G) = \mathcal{C}(G)+\mathcal{I}(G)$ counts the total number 
of $m$-cliques and $n$-independent sets in a graph $G$. For an $m$-vertex set 
$S_{\alpha} = \{ v_{1},\ldots ,v_{m}\}$, we define the edge set $E_{\alpha} = 
\{ e^{\alpha}_{k}: k=1,\ldots ,C(m,2)\}$ as the set of all edges connecting pairs 
of vertices $v_{i}, v_{j} \in S_{\alpha}$, and $C(m,2)$ is the number of ways of
choosing $2$ vertices out of $m$. If $S_{\alpha}$ corresponds to an $m$-clique 
in the graph $G$, the graph-string $g(G)$ must have $1$'s at all bit-positions 
associated with the edges of $E_{\alpha}$. Let the states $|0\rangle$ and
$|1\rangle$ satisfy $\sigma_{z}|a\rangle = (-1)^{a}
|a\rangle$. Then the operator $H_{\alpha} = \prod_{e\in E_{\alpha}} P_{1}^{e}$
(where $P_{1}^{e} = ( 1/2)\left[ I^{e} - \sigma_{z}^{e}\right]$, and $e$ labels
the qubit associated with edge $e$) will have $|g(G)\rangle$ as an eigenstate 
with eigenvalue $1$ when $S_{\alpha}$ is an $m$-clique, and zero otherwise. 
The operator that counts all $m$-cliques in a graph $G$ is then
$H_{cl}^{m} = \sum_{\alpha =1}^{C(N,m)} H_{\alpha}$, and by construction,
$H_{cl}^{m}|g(G)\rangle = \mathcal{C}(G)|g(G)\rangle$. A similar
analysis can be carried out for $n$-independent sets. Let $T_{\alpha} =
\{ v_{1}, \ldots , v_{n}\}$ be an arbitrary $n$-vertex set, and 
$\overline{E}_{\alpha}$ its corresponding edge set. If $T_{\alpha}$ is an 
$n$-independent set in a graph $G$, then the graph-string $g(G)$ must have
$0$'s at all bit-positions associated with the edges of 
$\overline{E}_{\alpha}$. The operator $\overline{H}_{\alpha} = 
\prod_{e\in\overline{E}_{\alpha}}P_{0}^{e}$ (where $P_{0}^{e}= ( 1/2)
\left[ I^{e}+\sigma_{z}^{e}\right]$, and $e$ labels the qubit associated with 
edge $e$) will have eigenstate $|g(G)\rangle$ with eigenvalue $1$ ($0$) 
when $T_{\alpha}$ is (is not) an $n$-independent set. The operator that 
counts all $n$-independent sets in an arbitrary graph $G$ is then
$H_{is}^{n} = \sum_{\alpha =1}^{C(N,n)} \overline{H}_{\alpha}$, and by
construction, $H_{is}^{n}|g(G)\rangle = \mathcal{I}(G)|g(G)\rangle$. For
calculation of $R(m,n)$, the problem Hamiltonian $H_{P}^{Nmn}$ is then
\begin{equation}
H_{P}^{Nmn} = H_{cl}^{m} + H_{is}^{n}.
\label{probHamltn}
\end{equation}
Note that $H_{P}^{Nmn}$ contains $\mathcal{O}(N^{s})$ terms, where
$N$ is the number of vertices and $s = \max\{ C(N,m),C(N,n)\}$. Since each 
$H_{\alpha}$ and $\overline{H}_{\alpha}$  is a projection operator, their 
operator norm will be unity and their matrix elements, being $0$'s and $1$'s, 
are specified with a single bit. Lastly, note that each term in
$H^{Nmn}_{P}$ is a product of at most $t=\max\{C(m,2),C(n,2)\}$
$\sigma_{z}$-operators so that $H_{P}^{Nmn}$ is a $t$-local 
Hamiltonian \cite{kit}. By using perturbative gadgets, it can be reduced to 
a $2$-local Hamiltonian \cite{kkr,OlTer,JoFar}. 

For a given Hamiltonian $H(t)$, two approaches have been demonstrated to 
experimentally implement AQE \cite{stef,peng,dwave}.  
Refs.~\cite{stef}, \cite{peng} partitioned the full evolution into 
$\mathcal{N}$ subintervals of duration $\Delta t = T/\mathcal{N}$ which 
are sufficently short that the propagator $U_{l}$ for each subinterval $l$ 
can be factored via a Trotter expansion. This approach was applied to 
three-qubit systems, though it can be used for arbitrary size qubit systems. 
Ref.~\cite{dwave} describes experiments using a quantum annealing device 
designed to implement adiabatic quantum optimization algorithms. Results are 
reported of AQE solution for the groundstate of randomly generated instances 
of an $8$-qubit quantum Ising spin glass. Work using perturbative gadgets is 
underway to convert $H^{Nmn}_{P}$ into a $2$-local form amenable to both 
AQE experimental approaches. 

\textbf{Ramsey Numbers and QMA:} Quantum complexity theory formalizes the 
notion of efficient quantum algorithms. Our interest is in the quantum 
complexity class \textit{QMA\/} which generalizes the randomized 
version of the classical complexity class \textit{NP} \cite{kit,kkr}.

\textit{QMA\/} is a class of promise problems where each problem $L$ is the
union of two disjoint sets of binary strings $L_{y}$ and $L_{n}$ corresponding
to Yes and No instances of the problem. For a string $x\in L_{y}\cup L_{n}$, 
the task is to determine whether $x\in L_{y}$ or $x\in L_{n}$ using  
polynomial resources. Let $\mathcal{H}$ denote a two-dimensional Hilbert 
space; and $|x\rangle$ the 
CBS labeled by the binary string $x$.
\begin{definition}[QMA]
Let $x\in L=L_{y}\cup L_{n}$ and $\epsilon = 2^{-\Omega (|x|)}$. The 
promise problem $L$ belongs to \textit{QMA} if there
exists a quantum polynomial-time verifier $V(|x\rangle ,|y\rangle) \rightarrow 
\{ 0,1\}$, and a polynomial $\pi (|x|)$ such that: (i)~for all $x\in L_{y}$, 
there exists an $|\xi\rangle\in\mathcal{H}^{\pi (|x|)}$ such that $Pr\{ 
V(|x\rangle ,|\xi\rangle )=1\} \geq 1-\epsilon$; and (ii)~for all $x\in L_{n}$ 
and $|\xi\rangle\in\mathcal{H}^{\pi (|x|)}$, $Pr\{ V(|x\rangle ,|\xi\rangle ) 
= 1\} \leq\epsilon$. Here $Pr\{ V(|x\rangle ,|\xi\rangle )=1\}$ is the 
probability that $V$ concludes $x\in L_{y}$ when the quantum witness is 
$|\xi\rangle$.
\end{definition}
Informally, if $x$ is a Yes (No) instance, there exists a (no) quantum witness 
$|\xi\rangle$ which causes $V$ to correctly (mistakenly) conclude $x\in L_{y}$ 
with probability at least $1-\epsilon$ (greater than $\epsilon$).

A promise problem is \textit{QMA\/}-Complete if it belongs to 
\textit{QMA\/} and all problems in \textit{QMA\/} are polynomially reducible 
to it. It has been shown \cite{kit,kkr} that $k$-\textit{Local Hamiltonian\/} is 
\textit{QMA\/}-Complete  for $k\geq 2$.
\begin{definition}[$k$-Local Hamiltonian]
Consider an $L$-qubit Hamiltonian $H=\sum_{j=1}^{r} H_{j}$, where $r=poly(L)$;
and each term $H_{j}$ acts on at most $k$ qubits ($k$-local); has operator norm 
$||H_{j}||\leq poly(L)$; and matrix elements specified by $poly(L)$ bits. 
Finally, two constants $a<b$ are specifed. The Hamiltonian
$H$ is a Yes instance if its groundstate energy $E_{gs}<a$, and a No instance
if $E_{gs}>b$. The problem is, given a $k$-local Hamiltonian $H$, determine 
whether $H$ is a Yes or a No instance.
\label{def2}
\end{definition}

Our Ramsey number AQE algorithm leads naturally to an example of $t$-Local 
Hamiltonian which we call RAMSEY. We have seen that the Ramsey problem 
Hamiltonian $H^{Nmn}_{P}$ is a $t$-local Hamiltonian; is a sum of a polynomial
number of terms $H_{j}=H_{\alpha}\;\mathrm{or}\; \overline{H}_{\alpha}$; and 
each $H_{j}$ satisfies the polynomial bounds specified in Definition~\ref{def2}.
Suitable choices for the constants $a$ and $b$ are $0.01<a<0.1$ and $b=1-a$.
Yes instances of RAMSEY then correspond to $N<R(m,n)$ since $E_{gs}=0<a$,
and No instances to $N\geq R(m,n)$ where $E_{gs}\geq 1>b$. It is possible to 
carry over the proof that $k$-Local Hamiltonian is in \textit{QMA\/} \cite{kit} 
to show that RAMSEY is also in \textit{QMA\/}. 

For an AQE algorithm with \textit{non-degenerate\/} ground-state (GS), the 
runtime is largely determined \cite{aqe} by the minimum energy gap $\Delta 
= \min_{t}\{E_{1}(t)- E_{0}(t)\}$. This connection fails for the Ramsey 
algorithm when $N=R(m,n)$ as the GS becomes \textit{degenerate\/} during its 
execution and so $\Delta$ vanishes. Determining how the runtime scales when 
$\Delta = 0$ (as with the Ramsey algorithm) is an open problem in adiabatic 
quantum computing.

In this paper we have presented a quantum algorithm that calculates 
two-color Ramsey numbers $R(m,n)$; numerically simulated the algorithm 
and shown it correctly determined small Ramsey numbers; discussed its 
experimental implementation; and shown that Ramsey number computation is 
in the quantum complexity class \textit{QMA}.

We thank W. G. Macready, P. Young, S. Jordan, and M. J. O'Hara for valuable 
comments, and F. G. thanks T. Howell III for continued support.


\begin{thebibliography}{99}
\bibitem{grah} R. L. Graham, B. L. Rothschild, J. H. Spencer, 
\textit{Ramsey Theory\/} (Wiley, New York, 1990).
\bibitem{hand} J. Ne\u{s}et\u{r}il, in \textit{Handbook of 
Combinatorics\/}, eds. R. L. Graham, M. Gr\"{o}tschel, L. Lov\'{a}sz 
(Elsevier, New York, 1995), Vol.~2, p.~1331.
\bibitem{boll} B. Bollob\'{a}s, \textit{Modern Graph Theory\/} (Springer, New 
York, 1998).
\bibitem{PHthm} J. Paris, L. Harrington, in \textit{Handbook of Mathematical
Logic\/}, ed.~J. Bairwise (North Holland, Amsterdam, 1977), p.~1133.
\bibitem{furst} H. Furstenberg, \textit{Recurrence in Ergodic Theory and
Combinatorial Number Theory\/} (Princeton, Princeton, NJ, 1981).
\bibitem{rosta} V. Rosta, \textit{Electronic J. Combinatorics\/} (2004), 
Dynamical Survey DS~13.
\bibitem{Ket&Sol} J. Ketonen and R. Solovay, Ann.\ Math.\ \textbf{113}, 
267 (1981).
\bibitem{aqe} E. Farhi, J. Goldstone, S. Gutmann, M. Sipser, 2000, available 
at arXiv.org:quant-ph/0001106v1.
\bibitem{spenc} J. Spencer, \textit{Ten Lectures on the Probabilistic Method\/},
2nd ed. (SIAM, Philadelphia, PA, 1994).
\bibitem{farhi} E. Farhi, J. Goldstone, S. Gutmann, J. Lapan, A. Lundgren, 
D. Preda, Science \textbf{292}, 472 (2001).
\bibitem{good} A. W. Goodman, Amer.\ Math.\ Monthly \textbf{66}, 778
(1959).
\bibitem{lhc} L. H. Clark, J. Graph Theory \textbf{16}, 451 (1992).
\bibitem{fg1} F. Gaitan, Int.\ J. Quantum Info.\ \textbf{4},  843 (2006).
\bibitem{fg2} F. Gaitan, Complexity \textbf{14}, issue~6, 21 (2009).
\bibitem{kit} A. Y. Kitaev, A. H. Shen, M. N. Vyalyi, \textit{Classical and
Quantum Computation\/} (American Mathematical Society, Providence, RI, 
2000), Section~$14$.
\bibitem{kkr} J. Kempe, A. Kitaev, O. Regev, SIAM J. Comput.\  \textbf{35},
1070 (2006).
\bibitem{OlTer} R. Oliveira, B. M. Terhal, Quant.\ Info.\ Comp.\  \textbf{8},
900 (2008).
\bibitem{JoFar} S. P. Jordan, E. Farhi, Phys.\ Rev.\ A \textbf{77}, 062329 
(2008). 
\bibitem{stef} M. Steffen, W. van Dam, T. Hogg, G. Breyta, I. Chuang,
Phys.\ Rev.\ Lett.\  \textbf{90}, 067903 (2003).
\bibitem{peng} X. Peng et al., Phys.\ Rev.\ Lett.\ \textbf{101}, 220405 (2008).
\bibitem{dwave} R. Harris et al., Phys.\ Rev.\ B \textbf{82}, 024511 (2010).
\end{thebibliography}
\end{document}